\documentclass[journal,comsoc]{IEEEtran}
\usepackage[T1]{fontenc}% optional T1 font encoding
\usepackage{color}
\usepackage{amsmath}
\usepackage{url}
\usepackage{array}
\usepackage{fancyhdr}
\usepackage{latexsym}
\usepackage{diagbox}
\usepackage{epsfig}
\usepackage{amssymb}
\usepackage{amsfonts}
\usepackage{amsxtra}
\usepackage{xspace}
\usepackage{makeidx}
\usepackage{graphics}
\usepackage{theorem}
\usepackage{subfigure}
\usepackage{algorithm,algpseudocode}
\usepackage{multirow}
\DeclareMathOperator*{\minimize}{minimize}

\begin{document}
	
	\title{Beamforming Design for Large-Scale Antenna Arrays Using Deep Learning}
	
	\author{Tian~Lin~\IEEEmembership{Student Member,~IEEE,} and Yu~Zhu,~\IEEEmembership{Member,~IEEE}
		\thanks{This work was supported by National Natural Science Foundation of China under Grant No. 61771147.}
		\thanks{T. Lin and Y. Zhu are with the Department of Communication Science and Engineering, Fudan University, Shanghai, China (e-mail: lint17@fudan.edu.cn, zhuyu@fudan.edu.cn).}}
	
	\maketitle
	
	\begin{abstract}
		Beamforming (BF) design for large-scale antenna arrays with limited radio frequency chains and the phase-shifter-based analog BF architecture, has been recognized as a key issue in millimeter wave communication systems. It becomes more challenging with imperfect channel state information (CSI). In this letter, we propose a deep learning based BF design approach and develop a BF neural network (BFNN) which can be trained to learn how to optimize the beamformer for maximizing the spectral efficiency with hardware limitation and imperfect CSI. Simulation results show that the proposed BFNN achieves significant performance improvement and strong robustness to imperfect CSI over the traditional BF algorithms.
	\end{abstract}
	
	\begin{IEEEkeywords}
		Deep learning (DL), millimeter wave (mmWave), beamforming (BF) design, large-scale antenna arrays, neural network (NN), beamforming neural network (BFNN).
	\end{IEEEkeywords}
	
	\IEEEpeerreviewmaketitle

	\section{Introduction}\label{sec:introduction}
	\subsection{Background and Motivations}\label{subsec:intro-A-background-motivation}
	Recently, hybrid analog and digital beamforming (HBF) design for millimeter wave (mmWave) communication systems with large-scale antenna arrays has been receiving much attention for its  advantage of providing high beamforming (BF) gains to compensate for the severe path loss at affordable hardware cost and power consumption \cite{2014Heath1,2014Heath2}. It has been recognized that 
	the most difficult part in the HBF optimization problem is the constant modulus constraint on analog BF due to its phase-shifter-based architecture \cite{LT2019}-\cite{2016Yuwei}. {\color{black}In the existing works using the model-based design approach to handle this difficulty, an orthogonal matching pursuit (OMP) based algorithm was proposed in \cite{2014Heath1}. However, the analog beamformer is limited to a pre-defined codebook. To enhance the performance of OMP, the manifold optimization method was applied in \cite{LT2019, 2016Zhangjun} for the analog BF optimization. An element wise iterative algorithm was proposed in \cite{2016Yuwei} to optimize the analog beamformer. However, all of these algorithms either require some approximations to simplify the original objective function, or a lot of serial time consuming iterations to obtain a solution. Moreover, perfect channel state information (CSI) is assumed in all of these algorithms.}
	
	In another aspect, recent works on intelligent communications have shown the great potential of the data-based deep learning (DL) method in dealing with the traditional challenging problems \cite{2018Ye}-\cite{2018Jinshi},\cite{2018He}. Inspired by these works, in this letter, we devote to applying the DL method to solve the complex BF design problem for mmWave systems with hardware limitation and imperfect CSI. Our motivations can be explained from the following three aspects:
	\begin{itemize}
		\item First, it is well known that as the BF design is quite a complicated non-convex problem due to the joint optimization of multiple variables and the constant modulus constraint, it is unlikely to find a closed-form optimal solution \cite{2014Heath1}. As DL has been regarded as an efficient method to deal with intractable problems \cite{goodfellow2016deep}, it would be interesting to see what could be obtained if using DL to solve the BF optimization problem.
		\item  Second, through a large number of training iterations with a lot of samples, the DL-based schemes have been shown to have the ability to understand the complicated characteristics of  wireless channels \cite{2018Ye}. Compared with the conventional works  assuming perfect CSI [1], [3]-[5], the DL based approach is expected to possess strong robustness to imperfect CSI.
		\item {\color{black}Most efficient traditional BF algorithms require time-consuming serial iterations with high-complexity \cite{2016Zhangjun, 2016Yuwei}. However, the neural network (NN) after offline training has low complexity with limited matrix multiplications and additions when deployed online. Besides, thanks to the acceleration of parallel computation, the DL-based schemes can operate fast and thus be more applicable for high speed communications.}
	\end{itemize}
	
	{\color{black}
		There have been some recent works on the DL-based HBF design [13][14]. However, the assumption of perfect CSI is made in [13], and the output beamformer does not directly meet the constant modulus constraint. In [14], the analog beamformer is limited to a pre-defined codebook, which normally leads to certain performance loss.}
	
	\subsection{Novelty and Contributions}\label{subsec:intro-B-contribution}
	In this letter, we propose a DL-based BF design approach and develop a BF neural network (BFNN) which can be trained to learn how to optimize the beamformer for maximizing the spectral efficiency (SE) with hardware limitation and imperfect CSI. The contributions can be summarized as follows:
	\begin{itemize}
		\item \textit{New design approach:} As the analog beamformer is implemented with analog phase shifters, we cannot follow the traditional full-digital design approach \cite{2018Ye,2017JSC} to replace the analog beamformer by a digital NN and train it in the whole communication link. Instead, we propose a novel DL-based design approach by developing the BFNN which directly outputs the optimized beamformer based on the input of the estimated CSI. 
		
		\item \textit{Novel Loss function:} Different from the traditional NNs, where the loss function is normally defined as the mean square error (MSE) between the transmitted symbols (labels) and the recovered ones \cite{2018Ye,2017JSC, 2018He}, in the BFNN, we propose an elegant loss function which is closely related to the performance of SE.
		
		\item \textit{Robustness to imperfect CSI:} {\color{black} We propose a two-stage design approach to make the BFNN robust to imperfect CSI. During the first offline training stage, the BFNN learns how to approach the ideal SE when it only has practical channel estimate as its input. By doing so, in the second online deployment stage, the BFNN can adapt itself to imperfect CSI and achieve robust performance to channel estimation errors.}
	\end{itemize}
	
	Since in the mmWave HBF design the optimal digital beamformer normally has a closed-form solution \cite{LT2019}-\cite{2016Yuwei}, as an initial work and for the easy of presentation, in this letter we focus on the analog BF design and consider the scenario of a large-scale antenna array with only one radio frequency (RF) chain. %However, after introducing the detailed design approach and showing the effectiveness through simulations, we will finally discuss the generality of the proposed BFNN to more complex scenarios.
	
	\section{System Model}\label{sec:systemmodel}
	\begin{figure}[t]
		\begin{center}
			\centering
			\includegraphics*[width=3.4in]{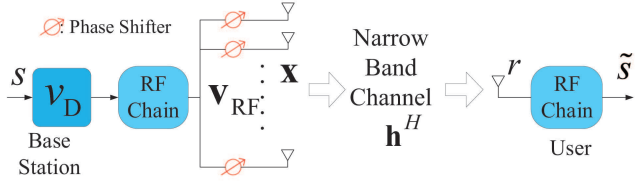}
			\caption{Diagram of  an MISO mmWave system with one RF chain.}
			\label{model}
		\end{center}
	\end{figure}
	Consider in Fig. \ref{model} the downlink of a narrowband multiple-input and single-output (MISO) mmWave system with the analog BF (precoding) architecture, where a base station (BS) with one RF chain and $N_\mathrm{t}$ antennas transmits one data stream to a user with only one receive antenna.\footnote{Although this letter focuses on the narrowband analog BF design with the aid of DL, the design approach can be generalized to a broadband MIMO mmWave system with HBF, which will be discussed in Section \ref{sec:discussion}.} Let $s$ denote the transmitted symbol with normalized average symbol energy, i.e., $\mathrm{E}\{|s|^2\}=1$. The symbol is first multiplied by a scalar digital precoder $v_\mathrm{D}$ (since there is only one RF chain, $v_\mathrm{D}$ is actually a scalar), and then by an $N_\mathrm{t}\times1$ analog precoder $\mathbf{v}_\mathrm{RF}$, which is implemented using phase shifters. The final precoded signal is then given by $\mathbf{x} = \mathbf{v}_\mathrm{RF}v_\mathrm{D}s.$
	%\begin{equation}\label{eqn:precoded-signal}
	%\mathbf{x} = \mathbf{v}_\mathrm{RF}v_\mathrm{D}s.
	%\end{equation}
	
	{\color{black}
		Through the MISO mmWave channel, the received signal at the user can be represented by $r=\mathbf{h}^H\mathbf{v}_\mathrm{RF}v_\mathrm{D}s + n$, where $n$ is the additive noise satisfying the circularly symmetric complex Gaussian distribution with zero mean and covariance $\sigma^2$, and $\mathbf{h}^H$ is the channel vector between the BS and the user. In this letter, a widely used Saleh-Valenzuela mmWave channel model [10][11] for $\mathbf{h}^H$ with one line-of-sight (LoS) path and $(L-1)$ non-LoS (NLoS) paths is adopted, which is  represented as 
		\begin{equation}\label{eqn:channel}
		\mathbf{h}^H=\sqrt{\frac{N_\mathrm{t}}{L}}\sum_{l=1}^L\alpha_l\mathbf{a}_\mathrm{t}^H\left(\phi_\mathrm{t}^l\right),
		\end{equation}
		where $\alpha_l$ denotes the complex gain of the $l$th path, and $\mathbf{a}_\mathrm{t}\left(\phi_\mathrm{t}^l\right)$ denotes the antenna array response vector at the BS, with $\phi_\mathrm{t}^l$ denoting the azimuth angle of departure associated with the $l$th path. In particular, the term with $l=1$ denotes the LoS component in $\mathbf{h}^H$.}
	
	In this letter, the SE, which has been widely used in the existing BF designs \cite{ 2016Zhangjun,2016Yuwei}, is chosen as the optimization objective, which is given as follows for the studied system
	\begin{equation}\label{eqn:SE-def}
	R = \mathrm{log}_2\left(1+\frac{1}{\sigma^2}\|\mathbf{h}^H\mathbf{v}_\mathrm{RF}v_\mathrm{D}\|^2\right).
	\end{equation}
	Considering the constant modulus constraint, $|[{\mathbf{v}}_{\mathrm{RF}}] _{i}|^2=1$, for $i=1,\ldots,N_\mathrm{t}$, and the maximum transmit power constraint $\|\mathbf{v}_\mathrm{RF}v_\mathrm{D}\|^2\le P$, it can be proved that the optimal $v_\mathrm{D}$ for maximizing $R$ is given by $\sqrt{P/N_\mathrm{t}}$. Then, the BF optimization problem for $\mathbf{v}_{\mathrm{RF}}$ is given by
	\begin{equation}\label{eqn:opt_prob}
	\begin{array}{cl}
	\displaystyle{\minimize_{{\mathbf{v}}_\mathrm{RF}}} &  \mathrm{log}_2(1+\frac{\gamma}{N_\mathrm{t}}\|\mathbf{h}^H\mathbf{v}_\mathrm{RF}\|^2) \\
	\mathrm{subject \; to} &  |[{\mathbf{v}}_{\mathrm{RF}}] _{i}|^2=1, \quad \mathrm{for}\:i=1,\ldots,N_\mathrm{t},%\forall i,
	\end{array}
	\end{equation}
	{\color{black} where $\gamma=\frac{P}{\sigma^2}$ denotes the signal to noise ratio (SNR). As the SNR can be generally considered to be more accurately estimated than the CSI, throughout this letter we assume $\gamma_\mathrm{est}=\gamma$, where $\gamma_\mathrm{est}$ denotes the estimated SNR, and focus on how to deal with imperfect CSI.}

	\section{DL Model and  Design of BFNN}\label{sec:deep-model}
	In this section, we first introduce the new challenges when applying the DL method to solve (\ref{eqn:opt_prob}) with imperfect CSI. Then, we describe in detail the BFNN architecture. {\color{black}Finally, we analyze its complexity.}
	
	%Although some conventional algorithms are applicable to the beamforming design problem in (\ref{eqn:opt_prob}), thanks to the advantages of DL technology, we devote to proposing a particular BFNN to solve (\ref{eqn:opt_prob}). Similar to \cite{2018Ye}, BFNN is trained offline and deployed online. However, the application of DL to beamforming design problem is not straightforward. In this section, we first introduce the challenges in solving by a NN. Then, we propose our consideration to solve these challenges. Finally, we briefly generalized it to the HBF case.

	\subsection{Challenges}\label{subsec:challenges}
	Since the analog beamformer has a specific architecture containing analog phase shifters, we cannot follow the traditional approach to replace it by a multi-layer NN and train it in the BS-user communication link \cite{2018Ye}-\cite{2018DLGao}. Here, we propose a different DL design approach by designing the BFNN that directly outputs $\mathbf{v}_\mathrm{RF}$ to solve (\ref{eqn:opt_prob}). However, this design is not trivial due to the following three challenges.
	\begin{itemize}
		\item \textit{What should the input of the BFNN be?} Most existing DL-based works \cite{2018Ye,2017JSC, 2018HBF} take the received baseband digital signal as the input. However, this approach cannot be applied here as the received signal itself is a function of the analog precoder to be optimized. Furthermore, the number of dimensions of the received signal is much lower than that of the precoder ($N_\mathrm{t}$) to be optimized.
		
		\item \textit{How to guarantee that the output ($\mathbf{v}_\mathrm{RF}$) satisfies the constraint?} As it is well known that complex output is not well supported by most DL frameworks (e.g., Tensorflow, pytorch), it would be more difficult to further impose the constant modulus constraint on the output.
		
		\item \textit{What should the label be when training the BFNN?} In almost all of the conventional intelligent communication designs \cite{2018DLGao, 2018He}, the label is set exactly as the transmitted bits or perfect CSI. However, it is difficult to find a proper label for this BF design problem. For example, if we take an optimized analog beamformer based on a traditional algorithm as the label,  the resulting BFNN would not perform better than the traditional algorithm.
	\end{itemize}
	
	\begin{figure*}[!t]
		\centering
		\includegraphics*[width=4.9in]{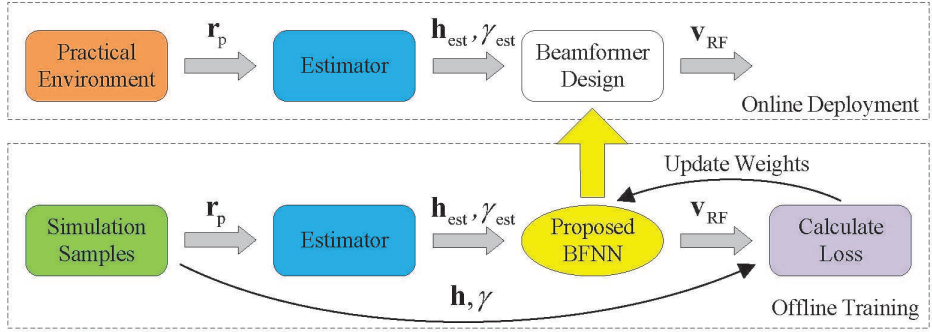}
		\caption{{\color{black}Illustration of the two-stage DL-based HBF design approach: offline training and online deployment.}}
		\label{fig:system}
	\end{figure*}
	
	% ================= Section III-B ==================
	\subsection{BFNN Architecture}\label{subsec:design}
	
	{\color{black}In this subsection we first elaborate on some considerations for the above three challenges, and then describe the two-stage design approach of the BFNN.}
	
	\subsubsection{{\color{black}Three Specific Considerations}}
	\begin{itemize}
		\item {\color{black} \textit{Input of the BFNN:} As the analog beamformer is implemented with analog devices, it cannot be replaced by a full-digital NN and trained in the whole communication link. Instead, the BFNN is designed to generate an optimized analog BF vector $\mathbf{v}_\mathrm{RF}$ based on the input of the channel estimate $\mathbf{h}_\mathrm{est}$ and the SNR estimate $\gamma_\mathrm{est}$.}
		%In most related works \cite{2014Heath1, LT2019, 2016Zhangjun, 2016Yuwei, 2019HBF}, perfect CSI is assumed available at the transmitter, which is unrealistic in practical systems. In this letter we assume a channel estimator is applied to estimate the CSI with limited feedback from user side, according to [2]. In order to  achieve robustness against channel estimation errors, we use the estimated CSI $\mathbf{h}_\mathrm{est}$  and $\gamma_\mathrm{est}$ as the input and the trained BFNN is supposed to achieve  robustness to  estimation errors when deployed online.}
		
		\item {\color{black}\textit{Lambda Layer:} To ensure that the output of the BFNN, i.e., $\mathbf{v}_\mathrm{RF}$, is a complex-valued vector satisfying the constant modulus constraint, a  self-defined Lambda layer is added at the end of the BFNN.} Specifically, letting $\boldsymbol{\theta}$ denote its real-valued input (the output of the last dense unit), its complex-valued output is given by
		\begin{equation}\label{eqn:Lambda-layer}
		\mathbf{v}_\mathrm{RF}=\exp(\mathrm{j}\cdot\boldsymbol{\theta})=\mathrm{cos}(\boldsymbol{\theta}) + \mathrm{j}\cdot\mathrm{sin}(\boldsymbol{\theta}),
		\end{equation}
		where $\mathrm{j}=\sqrt{-1}$. It can be seen from (\ref{eqn:Lambda-layer}) that $\boldsymbol{\theta}$ has a clear physical meaning that each element of $\boldsymbol{\theta}$ corresponds to the phase of each analog BF coefficient in $\mathbf{v}_\mathrm{RF}$. {\color{black} Compared with an alternative approach which is to first generate the real and imaginary parts of a complex value and then normalize that complex value on the unit circle, the proposed method directly generates the phase component, which automatically guarantees the constant modulus constraint with less neurons and makes the NN more elegant.} 
		
		\item \textit{Loss Function:} Different from traditional supervised learning designs \cite{2018Ye,2017JSC}, in our design, there is no need of labels and the BFNN is trained with the following new loss function directly related to the objective in (\ref{eqn:opt_prob})
		\begin{equation}\label{eqn:loss}
		{\mathrm{Loss}} = -\frac{1}{N}\sum_{n=1}^N \mathrm{log}_2(1+ \frac{\gamma_n}{N_\mathrm{t}}\|\mathbf{h}_n^H\mathbf{v}_{\mathrm{RF},n}\|^2),
		\end{equation}
		where $N$ denotes the total number of training samples, and $\gamma_n$, $\mathbf{h}_n$ and $\mathbf{v}_{\mathrm{RF},n}$ represent the SNR, CSI and output analog beamformer associated with the $n$th sample. Note that the reduction of the loss exactly corresponds to the increase of the average SE.
	\end{itemize}
	
	% ================= Overview of the BFNN =================
	{\color{black}\subsubsection{Two-stage Design Approach} With the above considerations on BFNN, the two-stage DL-based BF design approach is illustrated in Fig. \ref{fig:system}. In the  offline training stage, channel samples, transmit pilot symbols, noise samples are generated via simulation according to the system model in Section \ref{sec:systemmodel}. Then, a practical mmWave channel estimator is applied for the BS to obtain partial CSI. In this letter, we apply the classical mmWave channel estimator proposed in \cite{2014Heath2}, where the BS estimates the channel by sending pilot symbols with beamformers in a hierarchical codebook and receiving the feedback of the user's decision based on its received signal $\mathbf{r}_\mathrm{p}$. The channel estimate $\mathbf{h}_\mathrm{est}$ and the SNR estimate $\gamma_\mathrm{est}$ are sent into the BFNN as input. Note that as mentioned in Section \ref{sec:systemmodel}, it is assumed $\gamma_\mathrm{est}=\gamma$. Then, the BFNN generates the optimized analog beamforming vector $\mathbf{v}_{\mathrm{RF},n}$ by minimizing the loss function defined in (\ref{eqn:loss}). Since the channel samples and SNR values are generated via simulation, they (perfect CSI and SNR information) can be directly used in the calculation of the loss, as shown in Fig. \ref{fig:system}. As the DL-based technique is essentially a gradient-descend method, the loss can be guaranteed to converge to a local optimal with a proper learning rate [12]. By taking the channel estimate as the input of BFNN and using the perfect CSI in the loss function, the BFNN can be trained to learn how to approach the ideal SE with perfect CSI as much as possible and be robust to channel estimation errors.
		
		In the  online deployment stage, the same mmWave channel estimator is applied for the BS to obtain partial CSI. The channel estimate is then sent into the BFNN to obtain the optimized beamformer. It is worth noting that the perfect CSI is only required to compute the loss during the offline training stage. When deployed online, all parameters of the BFNN have already been fixed and the well-trained BFNN only accepts the imperfect CSI as the input and directly outputs the analog beamformer, with no need to take the perfect CSI to calculate the loss. }
	
	%Similar to \cite{2018Ye, 2018DLGao, 2018He}, $\mathbf{h}_n$ is generated randomly according to (\ref{eqn:channel}) via simulations, $\gamma_n$ is generated randomly between $-20\mathrm{dB}$ and $20\mathrm{dB}$. $\mathbf{v}_{\mathrm{RF},n}$ is the output of the BFNN based on the input of imperfect CSI $\mathbf{h}_{\mathrm{est},n}$ and $\gamma_{\mathrm{est},n }$. As the DL-based technique is essentially a gradient-descend method, the loss can converge to a local optimal with  proper learning rates [12].
	
	%{\color{black}We summarize the offline training and online deployment procedures in Fig. 2. Specifically, during the offline training, the simulated perfect CSI is estimated by the channel estimator to obtain imperfect CSI as the input, and then used to calculate the loss with the output analog beamformer. The loss can reduce iteratively (corresponding to the increase of the average SE) and the trained parameters will finally be fixed when the training is completed. Then for online deployment, the trained BFNN can accept imperfect CSI as the input and directly output the result $\mathbf{v}_\mathrm{RF}$ (perfect CSI is no longer required). With this loss function the BFNN is supposed to be trained to approach the ideal SE with perfect CSI as much as possible, and thus has certain robust performance to channel estimation errors when deployed online.}

	{\color{black} To show the detailed structure of the BFNN, we consider a MISO system  with $N_\mathrm{t}=64$ and show in Tab. \ref{Tab:BFNN} the network structure, the output dimension, the activation function (if used) and the number of trainable parameters of each layer. It will be shown in Section \ref{sec:simulation} that such a network can achieve excellent performance in the mmWave system setup there. As shown in Fig. \ref{fig:system}, since the input $\mathbf{h}_\mathrm{est}$ is a complex-valued vector and the BFNN is a real-valued network, the real and imaginary parts of $\mathbf{h}_\mathrm{est}$ are concatenated and further with $\gamma_\mathrm{est}$ to form a  $(2N_\mathrm{t}+1)\times 1$ real-valued input vector.  Three dense layers are then applied with 256, 128, 64 neurons, respectively. Similar to most related works [6]-[9][11][13][14], the number of dense layers and that of the neurons in each layer are determined by the empirical trials. Different activation functions (e.g., Relu, tanh and sigmoid) have been tried for the last dense layer, but simulation results suggested that the output without any activation function achieves the best performance and convergence behavior. To enhance the convergence, each dense layer is preceded by a batch normalization layer, which is omitted in the Tab. \ref{Tab:BFNN}. At the end, the Lambda layer imposes the constraint modulus constraint on the final output. To guarantee the generality of the BFNN, multiple samples are required for offline training. In our experiments, the training, validation, and testing sets contain $10^5$, $10^4$, and $10^4$ samples, respectively.}
	
	{\color{black}
		\subsection{Complexity Analysis}
		In this subsection we analyze the computational complexity in terms of the number of floating-point operations (FLOPs) for the proposed BFNN. As the complexity in the offline training stage is normally not counted [8], only that in the online deployment stage is counted. According to \cite{FLOPS}, the number of FLOPs of a dense layer is given by $(2N_\mathrm{I}-1)N_\mathrm{O}$, where $N_\mathrm{I}$ and $N_\mathrm{O}$ denote the input and output dimensions, respectively. In the case when $N_\mathrm{t}=64$, considering the BFNN in Tab. \ref{Tab:BFNN}, the total number of FLOPs of the BFNN is about 0.15 million. \par For  conventional model-based HBF designs [1][4][5],  the asymptotic computational complexity in terms of the number of complex multiplications is in the order of $\mathcal{O}\left(N_{\mathrm{t}}^{3}\right)$ as they involve the operations such as singular value decomposition and matrix inversion. Even taking the coefficient in the complexity order to be 1, the number of  $N_{\mathrm{t}}^{3}$ complex multiplications  is about 0.26 million with $N_\mathrm{t}=64$. It can be seen that the proposed BFNN has competitive computational complexity when compared with the traditional model-based HBF algorithms. In addition, the main operation of BFNN involves large-scale matrix multiplications and additions, which can be effectively accelerated by the graphics processing unit (GPU). However, most of the traditional model-based HBF algorithms normally involve serial iterations (the optimization of the next iteration depends on the result of the previous iteration) and are not suitable for parallel computing.}

	\section{Simulation Results}\label{sec:simulation}
	{\color{black} Throughout the simulations, a half-wave spaced uniform linear array with $N_\mathrm{t}=64$ is deployed at the BS. The Saleh-Valenzuela mmWave channel model in (\ref{eqn:channel}) with exactly the same parameters as those in \cite{2017Gao} is considered, where $L$ is set to 3, i,e, the channel contains one LoS path and two NLoS paths. $\alpha_l$ satisfies independently and circularly symmetric Gaussian distribution with zero mean, and the variance of $\alpha_l$ is set to $1$ for $l=1$, and set to $10^{-0.5}$ for $l=2,3$. $\phi_\mathrm{t}^l$ satisfies independently uniform distribution in $[-0.5\pi,0.5\pi]$. Two state-of-the-art HBF algorithms in the special case of one RF chain are considered for comparison, i.e., the manifold-optimization based HBF algorithm in \cite{2016Zhangjun} and the iterative HBF algorithm in \cite{2016Yuwei}. The classical channel estimation algorithm in \cite{2014Heath2} is applied for obtaining ${\mathbf{h}}_\mathrm{est}$. In the traditional HBF algorithms, $\mathbf{h}$ is directly replaced by ${\mathbf{h}}_\mathrm{est}$ when computing the BF coefficients. In the proposed BFNN, the hyper-parameter setting is shown in Tab. \ref{Tab:BFNN} and fixed throughout all experiments. The learning rate is initialized at 0.001 and the Adam optimizer is used. The real-valued loss function results in real-valued gradient of trainable parameters of the BFNN, and therefore can be directly implemented with Tensorflow. All source codes, data sets and some trained weights are provided openly in [15].}
	
	% \begin{figure}[t]
	% 	\centering
	%     \includegraphics*[width=2in, height=2.05in]{PNR.eps}
	% 	\caption{{\color{black}SE v.s. SNR for different BF algorithms with imperfect CSI of  different PNRs.}}
	% 	\label{fig:system}
	% \end{figure}

\begin{figure}[t]
		\centering
		\includegraphics[height=7cm,width=7.5cm]{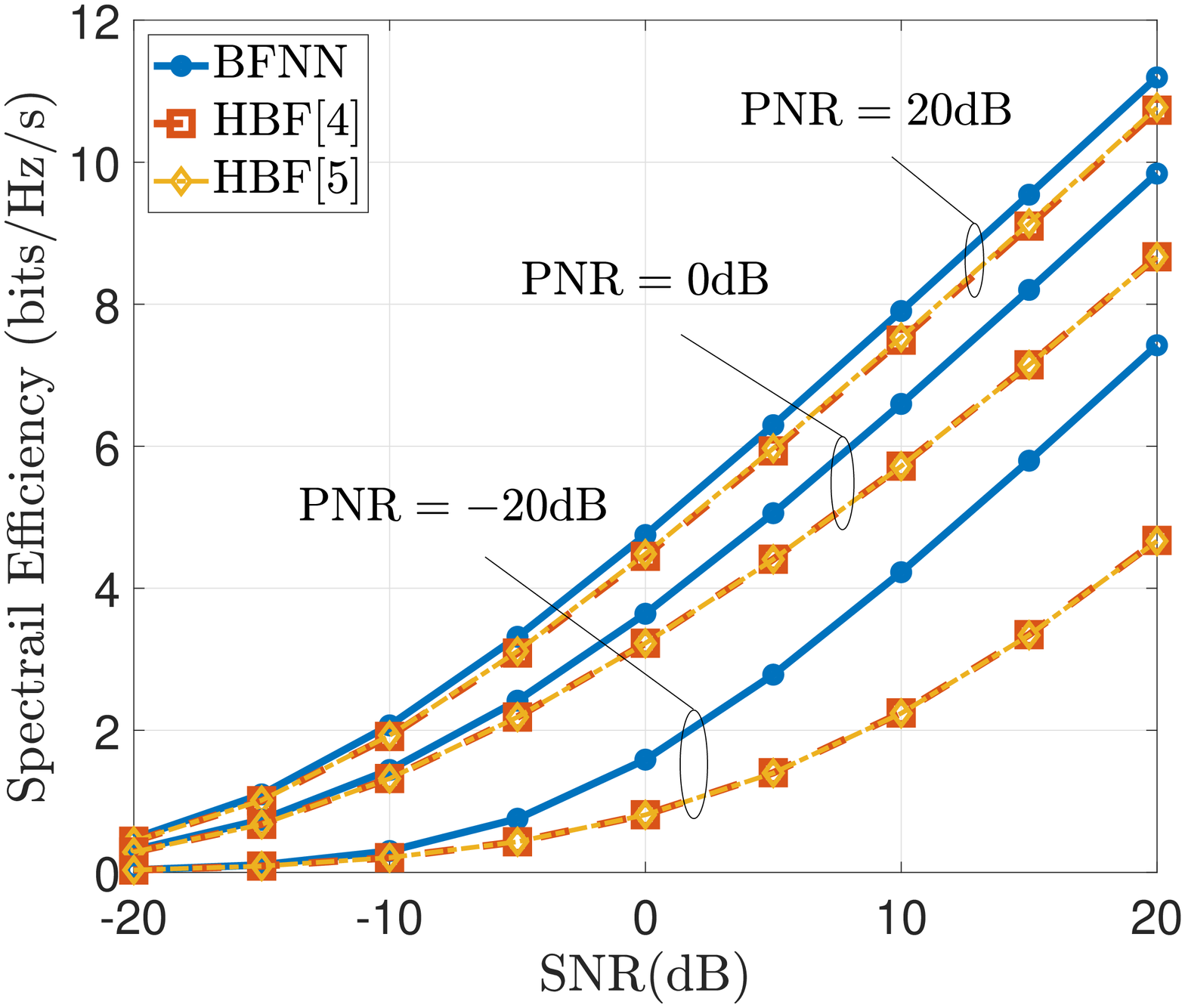}
		\caption{{\color{black}SE v.s. SNR for different BF algorithms with different PNRs. }}
		\label{fig:imperfect_CSI_PNR}
	\end{figure}

	{\color{black} Fig. \ref{fig:imperfect_CSI_PNR} shows the SE versus SNR performance under three channel estimation levels, which are characterized by three pilot-to-noise power ratios (PNRs), i.e., $-20\mathrm{dB}$, $0\mathrm{dB}$ and $20\mathrm{dB}$. It is assumed  that the estimation of the number of channel paths is correct, i.e., ${L}_\mathrm{est}=3$. The results in Fig. 3 show that with imperfect CSI the two traditional HBF algorithms perform quite similarly, but proposed BFNN significantly outperforms them. For example, at a SE of $8\mathrm{bits/s/Hz}$, the proposed BFNN achieves around a $1.5\mathrm{dB}$ gain in SNR over the traditional algorithms when $\mathrm{PNR}=20\mathrm{dB}$, and such a gain becomes larger for smaller PNRs.}
	
\begin{figure}[t]
		\centering
		\includegraphics[height=7cm,width=7.5cm]{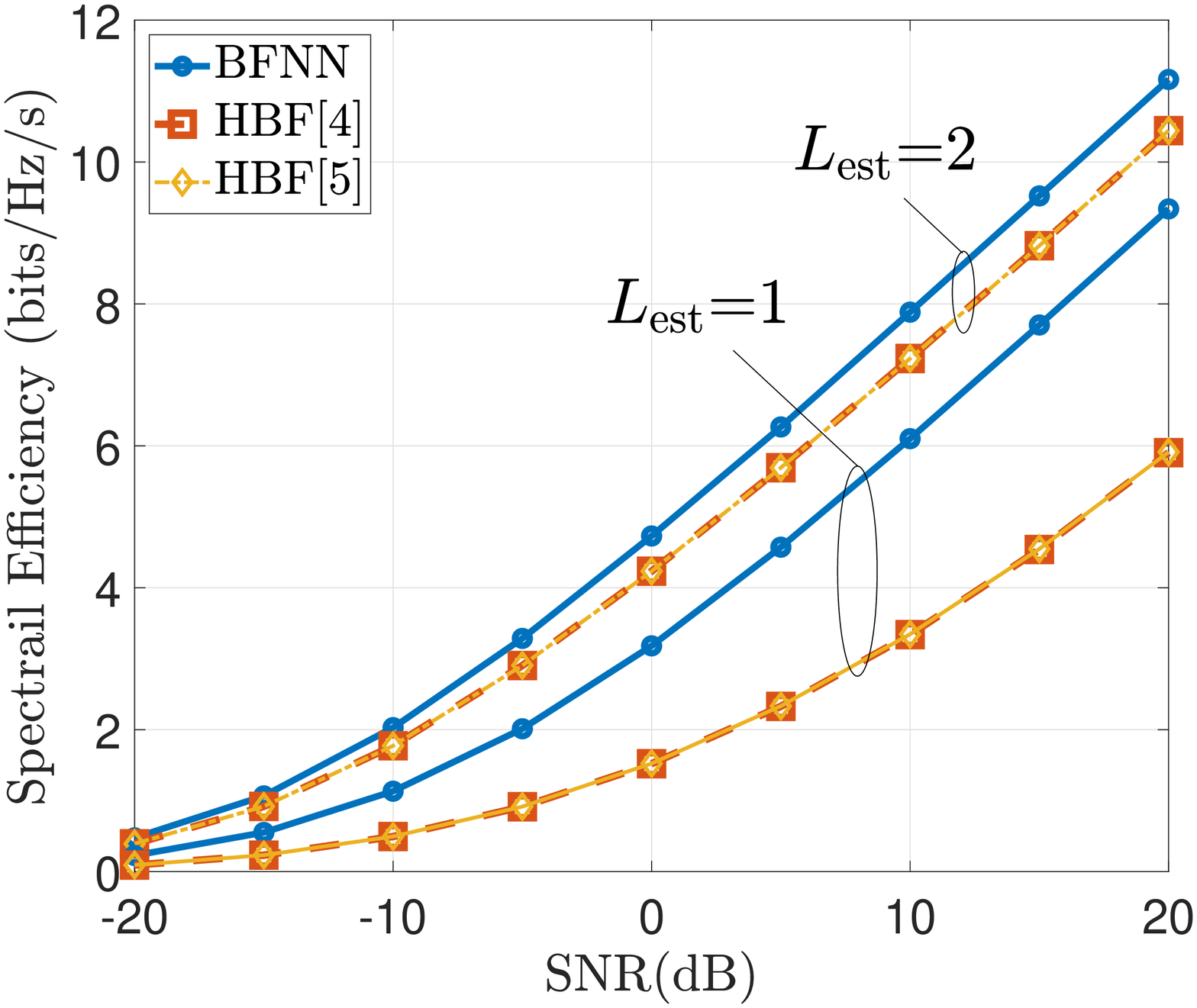}
		\caption{{\color{black}SE v.s. SNR for different BF algorithms with different $L_{\mathrm{est}}$. }}
		\label{fig:imperfect_CSI_Lest}
	\end{figure}

	In practical systems, there also exist estimation errors in estimating the number of total paths $L$. Due to the sparsity of the mmWave channels and considering the estimation complexity, $L_\mathrm{est}$ is preset to a small value \cite{2014Heath2}. Fig. \ref{fig:imperfect_CSI_Lest} shows the SE performance for different $L_\mathrm{est}$ values. It can be seen from this figure that the proposed BFNN outperforms the traditional HBF algorithms with the improvement becoming larger for less accurate $L_\mathrm{est}$.
	
	In summary, it can be concluded that the proposed BFNN exhibits much stronger robustness to imperfect CSI than the traditional algorithms. The less accurate the channel estimate is, the larger the performance gap is. This is because through many iterations with large training sets, the BFNN has been trained to learn the characteristics of the mmWave propagation channels, as well as the relationship between the imperfect CSI and the ideal SE with perfect CSI.
	
	%Conclusively, the simulation results in this subsection, suggest the DL-based scheme outperforms traditional beamforming schemes with imperfect CSI. In addition, it is also implied that under lower estimation accuracy, the proposed BFNN can achieve more improvement. The main reason of the results is that through many  iterations with large  training sets,  BFNN is trained to learn the characteristics of the mmWave channels and the relationship between the imperfect CSI and the SE (computed with perfect CSI). Therefore, compared with the conventional algorithms without corresponding design to counter estimation deviation, the proposed has stronger robustness with imperfect CSI.

	% \begin{figure}[t]
	% 	\centering
	% 	\subfigure[]{\includegraphics[height=3.8cm,width=4cm]{PNR_new.eps}}
	% 	\subfigure[]{\includegraphics[height=3.8cm,width=4cm]{Lest_new.eps}}
	% 	\caption{{\color{black}SE v.s. SNR for different BF algorithms with imperfect CSI. (a) With different PNRs. (b) With different $L_{\mathrm{est}}$. }}
	% 	\label{fig:imperfect_CSI}
	% \end{figure}

	% ================= Table 1 =====================
	\begin{table}[]
		{\color{black}
			\centering
			\caption{Implementation details of the BFNN.}\label{Tab:BFNN}
			\begin{tabular}{c|c|c|c}
				\hline
				Layer Name    & Output Dim. & Activation Func. & Number of  Paras. \\ \hline\hline
				Input Layer   & $129\times 1$        &   \diagbox{}{}         & 0             \\ \hline
				Dense Layer 1 & $256\times 1$       & ReLu       & 33024        \\ \hline
				Dense Layer 2 & $128\times 1$         & ReLu       & 32896        \\ \hline
				Dense Layer 3 & $64\times 1$          &  \diagbox{}{}           & 8256         \\ \hline
				Lambda Layer  & $64\times 1$          &  \diagbox{}{}           & 0             \\ \hline
		\end{tabular}}
	\end{table}
	% ============ End of  Table 1 =====================
	
	\section{Discussion of the Generality of BFNN} \label{sec:discussion}
	Although the BFNN is designed specifically for a simple scenario in this paper, it has good generality for other more complex problems. {\color{black}For example, in the broadband scenario, by concatenating the multi-tap channel vectors as the input and redefining the loss function to the one related to the broadband SE \cite{2016Zhangjun}, the BFNN can be extended to optimize the analog beamformer for broadband  mmWave channels.} For another example, considering the HBF problem with multiple RF chains,  while the optimal digital beamformer can be solved with a closed-form solution \cite{LT2019, 2016Yuwei}, a simple extension of the current BFNN is to increase the output dimensions from $N_\mathrm{t}$ to $N_\mathrm{RF} N_\mathrm{t}$ for the $N_\mathrm{t}\times N_\mathrm{RF}$ analog BF matrix with a new loss function. Following the similar idea, the BFNN can also be considered in the joint transmit and receive BF design or the multi-user BF design. %With optimized network structure and hyper-parameter tuning, it is supposed the DL-based scheme can show its advantages in these more general cases.
	
	\section{Conclusion and Future Work}\label{sec:conclusion}
	%We have proposed a DL-based BF design approach for mmWave systems with large-scale antenna arrays. With some special designs, the proposed BFNN can well handle the challenges of hardware limitation and imperfect CSI in mmWave systems. Although without extensive hyper-parameter tuning, simulations have shown the competitive performance of the BFNN and provided valuable insights for future BF designs. In our future work, it is of great interests to extend the original BFNN to more complex scenarios and investigate the corresponding physical meanings of each layers of BFNN.
	
	We have proposed a DL-based BF design approach for mmWave systems with large-scale antenna arrays. With some special designs on the self-defined Lambda layer and the loss function, the proposed BFNN can well handle the challenges of hardware limitation and imperfect CSI in mmWave systems. Simulation results have shown the competitive performance of the BFNN and provided valuable insights for future BF designs. 
	
	{\color{black}As for future work, due to the generality of BFNN, it is of interest to extend the BFNN to more complex BF problems. Furthermore, like that in the existing works [6][8][13][14], in this study the number of layers and that of the neurons in each layer in the proposed BFNN are mainly determined by the empirical trials, it is also of interest to investigate the physical meaning of each layer.} \\
	
	%We have proposed a DL-based BF design approach and developed the BFNN for mmWave systems with hardware limitation and imperfect CSI. By adding a self-defined Lambda layer, the BFNN directly outputs analog beamformer satisfying the constant modulus constraint. Taking the channel estimate as the input, the proposed BFNN is trained to approach the ideal SE with perfect CSI. Simulation results have shown that the BFNN achieves stronger robustness to imperfect CSI than the conventional BF algorithms.
	
	%\bibliographystyle{unsrt}
	%\bibliographystyle{IEEEtran}
	%\bibliography{LT}

\end{document}